\documentclass[twocolumn,aps,showpacs,preprintnumbers,amsmath,amssymb,amsfonts,showkeys]{revtex4}

\usepackage{graphicx}
\usepackage{epsfig}
\usepackage{dcolumn}
\usepackage{bm}
\usepackage{latexsym}

\begin{document}

\title{Molten-Salt Depleted-Uranium Reactor}


\author{ Bao-Guo \,Dong }  
\affiliation{\\
Department of Nuclear Physics, China Institute of Atomic Energy,
P.O. Box 275 (10), Beijing 102413, China }

\author{ Pei  \,Dong }  
\affiliation{\\
QianXiJiaYuan, Building 1 Unit 4 Room 601,Xin Town, FangShan District, Beijing 102413, China }

\author{ Ji-Yuan  \,Gu }  
\affiliation{\\
QianXiJiaYuan, Building 1 Unit 4 Room 601,Xin Town, FangShan District, Beijing 102413, China }

\date{\today}

\begin{abstract}
The supercritical, reactor core melting and nuclear fuel leaking accidents have troubled fission reactors for decades, and greatly limit their extensive applications. Now these troubles are still open. Here we first show a possible perfect reactor, Molten-Salt Depleted-Uranium Reactor which is no above accident trouble. We found this reactor could be realized in practical applications in terms of all of the scientific principle, principle of operation, technology, and engineering. Our results demonstrate how these reactors can possess and realize extraordinary excellent characteristics, no prompt critical, long-term safe and stable operation with negative feedback, closed uranium-plutonium cycle chain within the vessel, normal operation only with depleted-uranium, and depleted-uranium high burnup in reality, to realize with fission nuclear energy sufficiently satisfying humanity long-term energy resource needs, as well as thoroughly solve the challenges of nuclear criticality safety, uranium resource insufficiency and low-carbon development. They could provide safe, cheap, abundant, and clean energy resource and electric power lasting thousands years for humanity.
\end{abstract}



\maketitle


Energy resources are the principal foundation of economy and society developments. The requirement of energy resources due to world population increase and economic and social development is more and more. Fossil energy resource gross can be employed only enough to century and would bring abundant greenhouse gases while cause global climate to warm up. The density of renewable energy resources is too low, and industrial-scale exploitation would occupy a great deal of land resources, impact ecological environment etc. Simultaneously, it is very strongly limited by the windless, nonluminous, drought, and low-water from weather and climate conditions, hence it can only act as auxiliary energy resource. Otherwise, if human extracts a great deal of renewable energy resources from the earth environment, it consequently would impact and destroy the subtle balance of natural environment formed by long-term counteraction while bring unpredictable aftereffects. Only nuclear energy has very high energy density, cleanness and environmental conservation, without greenhouse gases trouble. If the fast neutron breeding fuel technology is adopted to make the most of uranium resource, its gross can supply the world to consume over thousands years. The gross of thorium in nature is 3--5 times that of uranium. Hence, only the gross of nuclear fission energy can supply all over the world to consume for thousands years.

The criticality safety, reactor core melting accident, and nuclear fuel leaking accident has troubled fission reactors for decades, and greatly limited their general applications. Fission reactors are already performed and operated all over the world. Otherwise, breeding reactors have been investigated for many years  \cite{DDG2,DDG4,DDG6,DDG7,DDG9}. However, these troubles are still open. So it is very important and urgent to originate innovative technology to solve the energy resource long-term supply challenge for the world.

The Molten-salt Depleted-uranium Reactor (MDR) is a reactor of with molten salts as primary coolant and only with its self-breeding nuclear fuel to realize long-term safe and stable operation after start-up  \cite{DDG12,DDG13}. Its nuclear fuel can include solid and liquid type. Its concept and principle of operation are suggested first. Here we only consider the research and development (R\&D) and application problem of a molten salt nuclear fuel type of reactor core.

　　MDR adopts uranium-plutonium cycle, fast-neutron spectrum, chloride molten salt, Depleted-Uranium (DU) and high burnup mode. It can adopt thorium-uranium cycle too. According to different operating temperature it can be divided into low, moderate and high temperature modes, working at about 600, 800 and 1000 ${^\circ}$C respectively. According to the reactor vessel shape it can be divided into spherical and cylindrical. According to the different application aim it can be divided into simple and general type. Simple type reactors are mainly for heat supply and industrial high temperature heat supply, general type mostly for power supply. Both can supply the combined heat and power too. Small type reactor can directly supply mechanical power as well. Simple cylinder type serves for fission product burner reactor, predominance distinctness.
MDR could realize with fission nuclear energy sufficiently satisfying humanity long-term energy resource needs, at the same time, thoroughly solve the challenges of nuclear criticality safety, uranium resource insufficiency and low-carbon development.



{\it MDR possesses extraordinary excellent characteristics}
(1) It will never explode, will not be prompt criticality or no supercritical accident, and ensure nuclear criticality safety.
(2) With negative feedback or component nuclear materials with self-stabilizing mechanism it can keep at critical state, to realize long-term safe and stable operation with negative feedback.
(3) In-reactor the whole uranium-plutonium cycle is finished, namely within the vessel forms the closed uranium-plutonium cycle chain. This can greatly shorten the industrial chain of nuclear power technology and reduce its complexity and difficulty, wholly optimize and greatly promote the quality of nuclear energy.
(4) When normal operation, it only needs with DU, itself has no need of uranium enrichment and fission product cleanup, and the optimization of extraction intermediate product protactinium, $^{233}$Pa, alone to decay, and solves the challenge of proliferation resistance entirely.
(5) It is capable of realizing the DU high burnup, and solves the problem of uranium resource insufficiency. At the same time, it can also realize the high transmutation of high dangerous fission products, and reduce the risk of ecological environment to suffer the radioactive pollution.

{\it MDR could overcome the disadvantages of the following reactors}
(1) Both solid nuclear fuel thermal-neutron reactors and sodium-cooled fast reactors have the existing risk to occur supercritical accident, reactor core melting and leakage of radioactive material accidents.
(2) The uranium resource utilization ratio of the thermal-neutron reactor is very low, and it has no means of everlasting in resource.
(3) Traditional sodium-cooled fast reactor, although it can breed nuclear fuel but it needs reprocessing to realize and perform the whole uranium-plutonium cycle and normal operation. Due to the high chemical activity, sodium makes its technology complication, costliness, and maintenance and repair difficulty and take long time, greatly reduce its energy resource values. Unless the sodium-cooled challenge could be solved, its energy resource value is not very high.
(4) The unique advantage of accelerator-driven system (ADS) is only its subcritical character. Its supercritical safety predominance is determined by its subcritical degree. But it has to add and match a piece of high-energy strong-current accelerator as cost. Otherwise, even if the reactor loads new nuclear fuel this reactor can not reach up to critical, thus, along with the burnup deepening its subcritical degree would get much deeper, that would need the outer neutron source stronger and stronger along with the burnup deepening. Withal the accelerator characteristic is difficult to withstand. Therefore it is difficult for ADS to apply actually.
(5) Most of thorium-based molten-salt reactors, due to its nuclear fuel neutron multiplication factor lower and transmutation period long, are thermal or epithermal neutron spectra. The breeding type requires on-line purge or cleanup and optimization, and all of the whole system is high radioactive, shielding and operation is difficult, hence it is greatly difficult to realize. Burning type needs to supply nuclear fuel and it is similar to the solid nuclear fuel thermal-neutron reactor.
(6) Traveling-wave reactor  \cite{DDG7} or Depleted-uranium reactor  \cite{DDG12} or CANDLE fast reactor  \cite{DDG6} has the similar principle, and adopts solid nuclear fuel and liquid metal coolant. Traveling-wave reactor and Depleted-uranium reactor mostly adopt the mature technology of sodium-cooled fast reactor, and CANDLE fast reactor adopts similar lead-cooled technology. Because the cladding life limitation can not reach up to its requirement of high burnup necessary condition, in the near future it is difficult to realize them. In addition the sodium-cooled challenge has to be improved and overcomed.

{\it MDR safety analysis}
(1) Criticality safety analyses. Because the supercritical safety of MDR is controlled and based on the physics law instead of exotic control, thus there is nonexistence control failure problem, by physical principle ensure no supercritical accidents, thoroughly solve nuclear criticality safety challenge already. This already makes it will never explode.
(2) Reactor core melting accident does not exist no longer for molten-salt cores or insignificance.
(3) Nuclear fuel leaking accident. When fuel salts directly leak out in air it would freeze and solidify whole radioactive materials except gas. Leak into other kind of molten salt to cause the change little; A little amount leaks into water would dissolve or deposit whereas slather leaks into water would cause steam explosion accident. But there is no water or no slather water in the MDR core.
(4) Emergency case, or when the molten salt temperature surpasses the set value of frozen salt overheating protection valve, molten salts in the vessel drains into geometrical subcritical storage tank, by natural cooling and the high heat capacity of molten salts absorbing, passively remove decay heat out. This is also the normal method to shutdown.
(5) Reactor core operates at low-pressure, without crack accident and/or related safety problem caused by interior high pressure.
(6) Molten salts directly exposes in air not to burn, only cooling and solidification.
(7) Molten-salt reactors have capacity for automatic following the change of output power. When no power output the reactor can auto-shutdown, thereby adding one safety protection function. They have a negative fuel molten salt void coefficient (expanded fuel is pushed out of the core) and negative thermal reactivity feedbacks, provide a kind of excellent passive nuclear safety characteristics.  The predominance of on-line charge consists in it could reduce nuclear fuel loading in the reactor core, to improve or avoid the fault of the criticality safety risk produced in the solid nuclear fuel reactor core due to nuclear fuel inventory over loading.

　　Comprehensive analysis results indicate that the integral safety of MDR is sky-high. The consequent results of the molten salt leaking accident are controllable and acceptable. It entirely could be built and operated within cities or industrial districts.

{\it MDR economy analysis}
(1) MDR core structure is simple. For instance, within the spherical vessel there are only molten salts. Contrarily, the more complicated the structure is, and the much larger the R\&D and constructional difficulty is, the much higher the corresponding cost too.
(2) It has no need of control rod and all of supporting facilities and Equipments for preventing supercritical accident.
(3) Reactor vessel operates at low pressure, as well no the risk of hydrogen chemical explosion in the water-cooling reactor, have no need of massive withstanding high-pressure explosion-proof equipment and vessel.
(4) Operating control is very simple. Realizable nobody watches automatic operation.
(5) Both DU and sodium chloride are abundant and low in price.
(6) It can realize high burnup of DU and high transmutation of fission products, reduce ecological shield cost.
(7) Basic construction requirement and cost are low.
(8) Without the restriction of breeding period, it is well suited for extensive generalization and application.

Overall, its construction and operation full cost is very low, thus the economy is very good and energy resource value is very high.

{\it Principle of operation of MDR}
　　The principle of operation of MDR is different from the traditional thermal-neutron reactor and sodium-cooled fast reactor at all. Both of the latter two reactors is burning the nuclear fuel stored within the reactor until to burn out then recharge. Because it stores abundant nuclear fuel within the reactor, far in excess of critical requirement, a great deal of redundant reactivity certainly results in that there is the risk of occurring supercritical accident or prompt critical. Hence, the supercritical accident risk is closely related to the principle of operation of the traditional thermal-neutron reactor and sodium-cooled fast reactor, and it has no means of thorough elimination.

MDR is according to the burning requirement to produce plutonium in the reactor core. It produces and burns plutonium at the same time, and it just produces what it needs plutonium and no more and no less.

Furthermore, how much it needs and how much it just produces, adjusted by its self-stabilizing mechanism, hence there is no redundant nuclear fuel within the reactor and nor redundant reactivity, and superadd that there is $\beta$-decay delay when nuclear fuel is produced. This point is crucial important. The time delay of producing plutonium makes the consumed plutonium cannot be supplied instantaneously. Hence, it makes that the chain reaction cannot develop on the explosion mode but it only can smoothly proceed. So it will never take place supercritical accident.

This could thoroughly solve the nuclear criticality safety problem. Hence if only this issue is considered it has very important significance for nuclear industry. Especially under the current situation people pay close attention to the nuclear safety.

　　When plutonium, $^{239}$Pu, absorbing a fast neutron fissions it releases the mean neutron number of 2.9. In all of the new generating neutron quantity one could make $^{238}$U absorb one neutron then by twice $\beta$-decay convert into $^{239}$Pu, and reuse one neutron to bring $^{239}$Pu to fission and release the next generation neutrons. There are a lot of residual neutrons to consume. From the point of view of neutron balancing namely critical, one only employs $^{238}$U to establish the continued chain reaction, it is entirely feasible to base on scientific principles.

　　By ignition, satisfying some definitely conditions, within the DU media in reactor, in the dynamical processes of producing and absorbing of neutrons, it builds a suitable neutron field, to realize the mutual promotion and limitation and compatible stable state of the self-sustaining breeding nuclear fuel cycle and nuclear fuel chain reaction. The self-stabilizing mechanism in this dynamical procedure can then realize the long-term stability of critical state. MDR is just based on such principle of operation of priority breeding then burnup and being capable of self-stabilization.

　　The self-stabilizing mechanism or negative feedback process of MDR is formed and originated by the following interdependent processes. To suppose sometime neutron flux suddenly increasing, the increase of the neutron flux results in plutonium faster burnup, this would instantaneously decrease the concentration of plutonium, then reduce the reaction rate of plutonium fission, and ultimately decrease the neutron flux to realize the negative feedback, and vice versa. Whereas new $^{239}$Pu nuclei should generate after about 3.3 days, namely that cannot instantaneously supply the consumption of plutonium. Comprehensive results are the system realizes the self-adjusting stable operation.

　　There are two methods to startup new reactors, one is to employ enriched uranium or plutonium to realize criticality ignition to start-up, the other is to employ the operating reactor to ignite and start-up new reactor with the entirely DU loading. Latter feasibility, solid fuel type CANDLE fast reactor already obtained theoretical calculation demonstration  \cite{DDG6} while molten salt fuel type awaits demonstration.

{\it MDR feasibility analysis}
　　First of all, the characteristic of the MDR structure is simple. One should not think the building complicacy and difficulty of the solid nuclear fuel thermal-neutron reactor and sodium-cooled fast reactor, whole or partly to apply mechanically to the R\&D of MDR, as that of MDR. This is ultimately improper on fundamentality. Because the most simple type reactor core structure of MDR, only requirement is to make a withdrawing low-pressure spherical vessel with stainless steel, charge in the compound molten salts produced according to special requirement, within the vessel there is nothing or no any structure at all except the salts, it can normally and stably operate in criticality safety state. This shows its degree of simplicity.

　　Secondly, the difficulty and key technology of MDR has been solved, include,

Within a molten salt type of DU media, we can establish up the sustained burning active zone or propagable nuclear burning waves. The key point of solution is to select and determine the plutonium concentration when reactor is critical. There is only a rather narrow plutonium concentration range that can realize it. If the plutonium concentration is too high or too low, the nuclear burning wave can not spread within the media. Plutonium concentration can be determined by calculations according to the specific reactor design requirement. Withal, both the traveling-wave reactor and CANDLE fast reactor, possessing similar principle of operation but different technology realization approach, are theoretically confirm already with MCNPX software etc by actual calculations, within the rectangular or columnar solid DU media, to build up the propagable nuclear burning wave, could be realized \cite{DDG7,DDG6}. Technically we have only to according to the special requirement to adjust the each component concentration of chloride molten salts, and it can satisfy the requirement of critical operation etc. UCl$_3$ + PuCl$_3$ + NaCl molten salts, may add MgCl$_2$ to reduce the melting point, can determine the suitable components to meet the needs, the technology is feasible, its property can satisfy the requirement of practical operation.

To keep the active zone or nuclear burning wave long-term stable technology. Primarily by the component of nuclear fuel and transmutable materials within reactor, build up negative feedback or self-stabilizing mechanism as stated in principle of operation, to realize self-stabilizing and keep critical state long-term stable.

To select and research and develop the materials which is compatible with chloride molten salts or corrosion resistance. Internationally the R\&D of molten-salt reactors has been for many years and makes great progress, and accumulates a great deal of mature technology as well. Its research results openly are provided already to international society  \cite{DDG9}. The research of molten salt, gas coolant and their compatible materials has been lasted for decades, at different operating temperature, the materials compatible with molten salt and gas coolant or corrosion resistance are already selected out. Typical compatible instance, with fluoride molten salts at and lower than 750 ${^\circ}$C, high nickel alloy, Hastelloy-N, at least can be employed for 30 years  \cite{DDG9}. Or by long-term in-depth research, there are already methods and corresponding projects, for instance, periodically replace parts corroded rather seriously. And this is not difficult. For operating temperature at and above 1000 ${^\circ}$C, materials are compatible with chloride molten salts or corrosion resistance, the first choice is TZM alloy or carbon-based composites \cite{DDG9}. That of middle and low temperature type may select and employ high nickel alloy, e.g. Hastelloy-N or stainless steels. All of these are confirmed by experimental examination, except not as good in-depth or long time as the experimental research and testing of fluoride molten salts. Experimental data indicate, without oxygen appearance, at suitable reactor operating temperature (under 600 ${^\circ}$C), no any applicative chloride-based molten salts, strongly corrode stainless steels or nickel-based alloy  \cite{DDG14}.

The technology realizes high burnup of DU and high transmutation of fission products and without fission product cleanup system. This is primarily realized by the innovative design of reactor core structure. For instance, experiment reactor optimized system is suggested first, i.e. simplified low operating temperature cylinder type. That is a cylindrical clapboard structure by horizontal disposal, and it could form the stable nuclear burning wave propagating along with the cylinder axis. The predominance of such kind reactor core structure consists in, without fission product cleanup system, not only can realize long-term safe and stable normal operation, but also can realize the high burnup of DU and high transmutation of high dangerous fission products. And it is well suited for industrial scale manufacture and applications  \cite{DDG13}.

　　Superadd, reactor core operation at low pressure, vessels no need of withstanding high-pressure, as well without requirement of other withstanding high-pressure equipment. The systemic technology outside reactor core is whole comparatively mature. On engineering, there have been many experimentally verified materials compatible with molten salts or gas coolants or corrosion resistance for selection. This shows that the integrated MDR system is feasible too.

　　In summary, MDR is feasible on all of the scientific principle, principle of operation, technology and engineering. At present this is the unique reactor which could make only burning DU become realization in the world. It could provide safe, cheap, abundant, clean energy resource and electric power lasting thousand years for humanity, realize with fission nuclear energy sufficiently satisfying humanity long-term energy resource needs, as well as thoroughly solve the challenges of nuclear criticality safety, uranium resource insufficiency and low-carbon development. Now we could affirm that MDR is an important breakthrough in the nuclear power technology field. It possesses the great potential to change the current world energy supply modes and patterns thoroughly.


\end{document}